\documentstyle[12pt]{article}
\def \slash#1{#1 \hskip -0.5em /}
\def \beq{\begin{equation}}
\def \eeq{\end{equation}}

\def \cfd{\Delta i_{23}}
\def \dfd{\Delta i_{26}}
\def \sma{\sigma^{\mu \alpha}}
\def \smr{\sigma^{\mu r}}
\def \sab{\sigma^{\alpha \beta}}
\def \nn{\noindent}
\input epsf
\title{\bf Is $B \to X_s \gamma$ equal to $b \to s \gamma$ ? \\
Spectator contributions to rare inclusive B decays. }
\author{{\bf John F. Donoghue, Alexey A. Petrov} \vspace*{5mm} \\
        Department of Physics and Astronomy \\
        University of Massachusetts \\
        Amherst, MA 01003 }
\date{}
\begin{document}
\maketitle
\vfill
\begin{abstract}
\nn
In $B \to X_s \gamma $ there are perturbative $QCD$ corrections at order
$\alpha_s$ not only to the single quark line process $ b \to s \gamma$
but also coming from a set of diagrams where the weak interaction vertex
involves a gluon which interacts with the spectator quark in the B hadron.
We discuss the impact of the full set of these diagrams.
These can influence the decay rate and also the line shape of the
photon spectrum as they favor a softer photon energy than does the
pure spectator decay $b \to s \gamma$. A subset of these diagrams generate
differences in the decay rates for charged and neutral B's , $B^+ \to
X_s \gamma$ vs. $B^0 \to X_s \gamma$ which can be searched for experimentally,
although our calculation indicates that the effect will be hard to
observe. Bound state effects turn out to be required to give rise
to spectator corrections to the decay rate at order $\alpha_s$.
The resulting contribution to the total rate is rather
small (of about $5 \% $ ), although the effect is somewhat larger on
portions of the photon spectrum. \\
$ $ \\
\bf{UMHEP-423}
\end{abstract}

\thispagestyle{empty}
\newpage
\setcounter{page}{1}

\section{Introduction}

Rare decays of heavy mesons received considerable interest in the recent
years as possible grounds for testing the Standard Model in flavor
changing neutral current processes which known to occur only at
one loop level \cite{hewett}. Exclusive decays of $B$-mesons to
the non-charmed
final states (which are examples of FCNC) are somewhat
easier to observe experimentally. However theoretical
calculations of exclusive rates include modeling of the
hadronization of the final
$s(d)$-quark and the resulting predictions vary by the order of magnitude
from one to another.
The inclusive radiative decay $B \to X_s \gamma$
was recently observed by CLEO collaboration
\cite{cleo} and is consistent with the $SM$ expectations.
It is now generally accepted that $QCD$ corrections significantly
change the decay rate for this process so it is important to bring
all of the possible corrections under control to
reduce theoretical uncertainties associated with $QCD$.

    Most work on the inclusive $B \to X_s \gamma $ decay proceeds by
calculating radiative corrections to the quark decay $b \to s \gamma$(see for
example \cite{quark}).
However, there are also several diagrams where a radiative correction involves
a gluon which interacts with the spectator quark. They generate effects
which are not accounted for in spectator model calculations, and
the inclusive rate $B \to X_s \gamma$ could in principle differ from
that of $b \to s \gamma$ at order $\alpha_s$. If one is interested in
a complete analysis of this decay mode, these non-spectator diagrams
must be included. This is especially true if one is interested in the
photon spectrum, since reactions which include the spectator quark produce a
softer photon spectrum than does the free quark decay.  The calculation
presented below
fills this gap and completes the calculation of the process of
inclusive radiative decay of heavy meson at the level
$O(\alpha_s)$

  In this paper, we present a $QCD$ based calculation which estimates these
non-spectator effects. There are 12 diagrams which we consider, shown
in Fig 1. The weak vertices are calculated in $QCD$ perturbation theory (some
vertices already exist in the literature), and the ACCMM model \cite{accmm}
is used to
calculate the decay rates and spectra. The interference of these diagrams
with the leading $b \to s \gamma$ amplitude only occurs for small values
of the momentum of the spectator quark. This portion is then particularly
sensitive to the bound-state wavefunction, but fortunately the interference
is relatively small. The main non-spectator effect comes from the region
of phase space where the spectator quark recoils significantly, and hence
are second order in $\alpha_s$.

The plan of the paper is as follows. In Section 2, we organize the calculation
and recall some of the information which in known about the gluonic vertices.
Section 3 is devoted to most of the details about the calculation of the
modifications to the decay rate and photon spectrum. The impact of the results
is described in Section 4. Before proceeding, let us resolve a nomenclature
ambiguity. It is conventional to call calculations which focus on the
$b \to s \gamma$ transition by the phrase "spectator model" and to refer to the
light quark in the B hadron as the "spectator" quark. We will continue to
use the latter phrase for the light quark in the B meson although, in the
diagrams which we consider, it is no longer a spectator but is an active
participant in the weak decay. We will call the $b \to s\gamma$ diagrams
the spectator process and conversely will label those which actively involve
the
spectator quark as non-spectator transitions.

\section{Diagrams and vertices}

There are quite a number of diagrams at order $\alpha_s$ which lead to
modifications to $B \to X_s \gamma$, as shown in Fig 1. Let us
divide these into five classes:

1. $QCD$-corrections to the single quark process $b \to s \gamma$.

2. ``Gluonic-penguin'' (bremstrahlung) diagrams which include $bsg$-vertex with
photons emitted from external quark legs.

3. ``Triangle'' diagrams in which both the gluon and photon are emitted
from weak vertex due to a quark loop.

4. Hard gluon spectator exchange diagrams with gluons emitted from $b$ or
$s$-lines.

5. Diagram ($10$) involves the $WW \gamma$-vertex and and is suppressed
by a power of $1/M_{W}^{2}$. We will not consider this diagram further.


The calculation proceeds by evaluating each of the diagrams, adding them
together coherently, integrating over the initial particles momentum using
a momentum space wavefunction and summing over the final momenta and spins.
The effective vertices are often momentum dependent and depend on the momentum
flow in the diagram. Since we are dealing with inclusive decays, the
details of hadronization of the final $s$-quark state are not relevant
for our discussion (all hadronic final states are summed over), so we
have to model only the initial state. This substantially reduces
the model-dependence of the final result. However we do have to
carefully account for interference effects between the many diagrams, as
different diagrams contribute to different portions of the final particles'
phase space. For the $b \to s \gamma$ process, the spectator quark emerges
with only low momentum reflecting its momentum in the bound state wavefunction.
However in the other diagrams, the final quarks carry a typical momentum of
a fraction of the B mass. Interference between the two classes only occurs
in a portion of the available phase space. We address this region specifically
below. For the initial state wavefunction we use the ACCMM model.

In this calculation we shall encounter two types of vertices: ``penguin'' and
``triangle''.

1.''Penguin'' vertex (Fig.2a). This vertex was calculated quite
a while ago \cite{shifman} and for the $bsg$-interactions turned
out to be
\beq \label{penguin}
\Gamma_{\mu}^a=-\frac{G_F}{\sqrt{2}} V_{bt} V_{ts} \frac{\lambda^a}{2}
\frac{g_s}{8 \pi^2} \Bigl( F_1^g (x) ( k^2 \gamma_\mu - k_\mu \slash{k} )
(1 + \gamma_5) - m_b F_2^g (x) i \sigma_{\mu \nu} k^\nu (1-\gamma_5) \Bigr)
\eeq
where $x = m_i^2/m_W^2$ for $i=c,t$. $F_1(x),F_2(x)$ are the Inami-Lim
functions, $F_i^g=\sum_{k} F_{ik}^g V_{bk} V_{ks}$, using unitarity
of CKM matrix we get
$F^g_1=-5.24$ and $F^g_2=-0.19$ for $m_t=176~GeV$,$m_c=1.3~GeV$.
Only the magnetic-moment type vertex contributes to the
real photon emission in the $b \to s \gamma$, so
\beq \label{onshell}
\Gamma_\mu = \frac{G_F}{\sqrt{2}} \frac{e Q}{8 \pi^2} V_{bi}V_{is}
m_b F_2^{\gamma} (x) i \sigma_{\mu \nu} q^\nu (1-\gamma_5)
\eeq
$F_2^{\gamma}(x) \approx 0.65$. We refer to $k$ being the
momentum of the gluon and $q$ - momentum of the photon. In the gluon
vertex, we shall drop the small contribution from $F_2^g$, which,
however, might be sizable for gluonic penguin-mediated hadronic
processes \cite{deshpande}.

2. ``Triangle'' vertex (Fig.2b). The calculation of this vertex reduces
to the calculation
of the set of usual triangle diagrams with certain form-factors ( a complete
parameterization of the diagrams of this type was first given in \cite{rosen})
and is interesting in the sense that light (spectator quark) current
couples to $b \to s$ current through $\epsilon$-tensors. The vertex itself
was recently calculated by D.Wyler and H.Simma in \cite{wyler} and takes the
form
\beq \label{triangle}
\Gamma_{\mu \nu}^a=-\frac{G_F}{\sqrt{2}} \frac{g_s e Q}{8 \pi^2}
\frac{\lambda^a}{2} V_{bi} V_{is}
\Bigl ( i f_{1~\mu \nu s} + i f_{2~\mu \nu s} +
i f_{3~\mu \nu s} \Bigr ) \gamma^s ( 1 + \gamma_5 )
\eeq
where
\begin{eqnarray}
f_{1~ \mu \nu s} =  \epsilon_{r \mu \nu s} (k^r \Delta i_5 + q^r
\Delta i_6) = \epsilon_{r \mu \nu s} l^r \nonumber \\
f_{2~ \mu \nu s} =  \frac{\epsilon_{\rho \sigma \mu s}}{kq}
 k^\rho q^\sigma k_\nu \Delta i_{23} \\
f_{3~ \mu \nu s} =  \frac{\epsilon_{\rho \sigma \nu s}}{kq}
k^\rho q^\sigma q_\mu \Delta i_{26} \nonumber
\end{eqnarray}
with
\begin{eqnarray}
\Delta i_5=-1+\frac{z_1}{z_0-z_1}(Q_0(z_0)-Q_0(z_1))-\frac{2}{z_0-z_1}
(Q_-(z_0)-Q_-(z_1)) \nonumber \\
\Delta i_6=1+\frac{z_1}{z_0-z_1} (Q_0(z_0)-Q_0(z_1))+\frac{2}{z_0-z_1}
(Q_-(z_0)-Q_-(z_1)) \nonumber \\
\Delta i_{23}=\Delta i_5 = - \Delta i_{26} \\
Q_-(x)=\int_0^1 \frac{du}{u}~ln(1-xu(1-u)) \nonumber \\
Q_0(x)=\int_0^1 du~ln(1-xu(1-u)) \nonumber
\end{eqnarray}
and $z_0=s/m_i^2$,$z_1=k^2/m_i^2$, $s$ is a total momentum of internal
quarks.

These vertices will be used to calculate the rate and spectrum in the next
section. We define the total decay rate as
\beq
\Gamma_{tot}=\Gamma_{q}+\Gamma_{sc}+\Gamma_{mix}
\eeq
where $\Gamma_{q}$ represents the leading contributions from $b \to s \gamma$
(diagrams of the class $I$), $\Gamma_{sc}$ gives the corrections which
involve the spectator quarks, and
$\Gamma_{mix}$ gives the mixing of the two. In powers of the
strong coupling constant, $\Gamma_{mix}$ is of order $\alpha_s$ and
$\Gamma_{sc}$ is of order $\alpha_s^2$.


In order to account for bound state effects, we will treat the decay of
a $B$-meson as a  $2 \to 3$ scattering process, assuming that the scattering
occurs from the bound state and model this state with the
ACCMM-model \cite{accmm}.
In calculating the decay rate (cross section $2 \to 3$)
we average over the spins of the initial particles which, in principle, leads
to the prediction for $\Gamma_{tot}$ for the ``mixture'' of $B$ and
$B^*$-particles (one can, at the expense of a considerable increase in
complexity of the calculation, separate $B$ and $B^*$
by introducing polarization operators for projecting out different
spin combinations). Thus the decay rate is defined as
\beq \label{rate}
d\Gamma=\frac{1}{4} ~\frac{1}{4 p_b p_u}~ |T_{fi}|^2~ d \Phi
\eeq
where factor of $4$ in the denominator comes from the spin averaging and
\beq \label{phase}
d\Phi=(2\pi)^4\delta^4(p_f-p_i) \frac{d^3p_s}{(2\pi)^32E_s}
\frac{d^3 p_u'}{(2 \pi)^32E_u'} \frac{d^3q}{(2 \pi)^3 2\omega}
\eeq
We also treat spectator quark as massless to simplify
the following calculations. The final phase space integration is performed
numerically.

\section{Spectator Contributions to the Rate and the Shape of the
Photon Spectrum.}

\subsection{Corrections of order $\alpha_s$.}
The introduction of interactions with the
light quark obviously makes the kinematics of the
process more involved - the final state is no longer a simple two-body
final state which results in delta-function-like peak in the
photon energy distribution. Instead, the peak is now extended down to lower
photon energies making it
possible for the photon to populate the energy interval from $0$ to
$ \sim~2.5~GeV$. Moreover, a difference between the decay spectra of neutral
and charged mesons also arises.

The question that naturally appears here
is whether it is possible to have an interference
of the spectator and non-spectator mechanisms. Since the $b \to s \gamma$
transition is of order $\alpha_s^0$ and the non-spectator matrix elements
are first order in $\alpha_s$ this would provide
corrections of order $\alpha_s$. If there were not any momentum
associated with the bound state, the spectator quark would remain
precisely at rest in the $b \to s \gamma$ transition and therefore would not
interfere with the diagrams where momentum was transferred to the spectator
quark. However if bound state effects are present the spectator quark will
always carry some momentum (typically of order 1 GeV or less) and
the two classes of diagrams can interfere. Since this occurs only over one
portion of the phase space, we can say that the interference effects are
really of order $\alpha_s 1GeV/m_B$.

In the bound state, the initial momenta are described by a momentum
distribution $\phi_B(p)$ (which is given by the wave function of quarks
inside of the meson).The interference takes place with a probability
proportional to a wave-function overlap. We now turn to the calculation
of this overlap.

Let $O_I$ to be an operator responsible for the transitions via
mechanism $I$ and $O_{II}$ - via mechanisms $II-IV$. Then, the mixing term
arises as a result of
\begin{eqnarray}
|M_{fi}|^2=\sum_{X_s} \langle B'|O_I+O_{II}|X_s \rangle \langle X_s
|O_I^+ + O_{II}^+|B \rangle = \nonumber \\
= \langle B'|(O_I+O_{II})(O_I^++O_{II}^+)|B \rangle =
 \langle B'||O_I|^2+|O_{II}|^2+2 Re~ O_IO_{II}^+|B \rangle
\end{eqnarray}
where we have used completeness relation for $|X_s \rangle$-states.
We are interested in the last term.

Let us write down the contributions from different classes of diagrams in turn.

\begin{eqnarray} \label{t2}
T_{II} = - \frac{g_s e}{8 \pi} \frac{1}{k^2}
\Bigl \{ \Bigl( Q_{b} \Bigl(
\frac{p_b \epsilon}{p_b q} - \frac{p_s \epsilon}{p_s q} \Bigr)
+ Q_{sp} \Bigl( \frac{p_u \epsilon}{p_u q} -
\frac{p_u' \epsilon}{p_u'q} \Bigr) \Bigr)
\bar{s} (p_s) \Gamma_\mu^a b (p_b) \cdot \nonumber \\
 \bar{u} (p_u') \gamma^\mu \lambda_a u (p_u) +
\bar{s} (p_s) \Bigl( \Gamma_\mu^a \slash{q} \frac{\slash{\epsilon}}{2 p_b q}
+ \frac{\slash{\epsilon}}{2 p_s q} \slash{q} \Gamma_\mu^a \Bigr)
b (p_b)~ \bar{u} (p_u') \gamma^\mu \lambda_a u (p_u) +  \\
\bar{s} (p_s) \Gamma_\mu^a b (p_b) ~ \bar{u} (p_u')
\Bigl( \gamma^\mu \frac{ \slash{q} }{2 p_u q} \slash{ \epsilon } +
\slash{\epsilon} \frac{ \slash{q} }{2 p_u' q} \gamma_\mu \Bigl) \lambda_a
u (p_u) \Bigr\} \nonumber
\end{eqnarray}
Here $\epsilon$ is the polarization vector of the photon
$Q_b, Q_{sp}$ are the charges of b-quark and spectator respectively,
and $\Gamma_\mu^a$ is given by (\ref{penguin}).
Note that $k^2$ in the denominator due to gluon propagator is to be cancelled
by $k^2$ in $bsg$-vertex, so $T_{II}$ is a local operator. This operator
(convoluted with the lowest order non-spectator graph, along with its
charge conjugate part) gives a contribution to the
shape of the photon spectrum.

The result of convolution can be represented as

\begin{eqnarray}
T^2_{II}=\frac{G_F^2}{8 \pi^2} |V_{bt} V_{ts}|^2 F_1^g (x) F_2^\gamma (x)
\alpha_s (m_b) \alpha  m_b C^1_\alpha \nonumber \\
\Bigl\{~p_s^\nu A^\alpha q^\beta
\bar{b} (p_b) \gamma_\mu \gamma_\nu \sigma_{\alpha \beta} ( 1 - \gamma_5 )
b (p_b') ~ L^\mu + \nonumber \\
+ \frac{Q_b^2}{2 p_sq~p_bq} \Bigl( S_b ( q_\mu g_{\alpha \nu} - g_{\mu \alpha}
q_\nu ) p_{s~\theta} - i P_b \epsilon_{\mu \alpha t \nu} q_t p_{s~\theta}
\Bigr) q^\beta~ \\
\bar{b} (p_b) \gamma_\nu \gamma_\theta
\sigma_{\alpha \beta} ( 1 - \gamma_5 ) b (p_b')~L^\mu +
\frac{Q_{sp}^2}{2 p_u'q~p_uq} p_s^\nu q^\beta~\bar{b}
(p_b) \gamma_\mu \gamma_\nu
\sigma_{\alpha \beta} ( 1 - \gamma_5 ) b (p_b') \nonumber \\
\Bigl( S_u(q_\mu g_{\alpha \theta} -
g_{\mu \alpha}q_\theta) ~ L^\theta -
i P_u \epsilon_{\mu \alpha t u} q_t ~L_5^u \Bigr) \Bigr\} + ~C.c. \nonumber
\end{eqnarray}
with $S_b=p_sq + p_bq$, $P_b=p_bq-p_sq$ and analogously for $S_u$ and $P_u$,
$L^\mu = \bar{u} (p_u') \gamma^\mu u (p_u)$,
$L_5^\mu = \bar{u} (p_u') \gamma^\mu \gamma_5 u (p_u)$, $C^1_\alpha=4/3$ is a
color factor coming from the averaging over initial colors and summing
over the final ones (taking into
account the fact that initial and final states are color-singlets).
Also
\begin{eqnarray}
A^\mu = \frac{p_b^\mu}{p_b q} +
\frac{p_u^\mu}{p_u q} - \frac{p_s^\mu}{p_s q} -
\frac{{p'}_u^\mu }{p_u'q}, \nonumber \\
\bar{b} (p_b) \gamma_\mu \gamma_\nu \sigma_{\alpha \beta} ( 1 - \gamma_5 )
b (p_b') = (g_{\mu \nu} g_{\alpha \beta} +  \nonumber \\
g_{\nu \alpha} g_{\mu \beta} -
g_{\mu \alpha} g_{\nu \beta} - i \epsilon_{\mu \nu \alpha \beta} )
\bar{b} (p_b') ( 1 - \gamma_5 ) b (p_b) - \\
i \bar{b} (p_b') ( g_{\mu \nu} \sigma_{\alpha \beta} + g_{\nu \alpha}
\sigma_{\mu \beta} - g_{\mu \alpha} \sigma_{\nu \beta} -
i \epsilon_{\mu \nu \alpha \theta} \sigma_{\theta \beta} )
( 1 - \gamma_5 ) b (p_b) \nonumber
\end{eqnarray}
The calculation of the contribution from the diagrams of the class
$III$ is more involved mainly because of the complicated structure
of the vertex (\ref{triangle}). The amplitude squared reads

\beq
T_{III}^2=\frac{1}{k^2} H_\mu^{\alpha \beta} L^\mu W_{\alpha \beta}=
-\frac{Const}{k^2}h^{\alpha \alpha}_\mu l^\mu
\eeq
with
\begin{eqnarray} \label{de}
Const= - \frac{G_F^2}{8 \pi^2} m_b |V_{ts}V_{tb}|^2 \alpha Q^2 \alpha_s
C^1_\alpha F_2^\gamma (x) \nonumber \\
l^\mu = \bar{u}(p_u') \gamma^\mu u(p_u) \\
h_\mu=2 \bar{b}(p') \sum_i f_{i \mu \alpha s} \sigma_{\alpha \nu} q^\nu
\slash{p} \gamma^s (1+\gamma_5) b(p) \nonumber
\end{eqnarray}
We will be working explicitly with components of this current
in order to get $|T_{III}|^2$. Using various $\gamma$-matrix
identities
\begin{eqnarray}
\sigma_{\alpha \nu}=\frac{i}{2} \Bigl[ \gamma_\alpha;\gamma_\nu \Bigr],
\nonumber \\
\gamma_\alpha \gamma_\nu \gamma_\theta = g_{\alpha \nu} \gamma_\theta +
g_{\nu \theta} \gamma_\alpha - g_{\alpha \theta} \gamma_\nu -
i \epsilon_{\alpha \nu \theta \beta} \gamma_\beta \gamma_5
\end{eqnarray}
we arrive at
\beq
h^\mu=2i q^\nu p^\theta \bar{b}(p') \sum_i f^\mu_{i \alpha s}
\Bigl\{g_{\nu \theta} \gamma_\alpha \gamma_s - g_{\alpha \theta}
\gamma_\nu \gamma_s + i \epsilon_{\alpha \nu \theta \beta} \gamma_\beta
\gamma_s \Bigr\} (1+\gamma_5) b(p)
\eeq
Note that the term symmetric under $\alpha \leftrightarrow \nu$ disappears.
Next, using explicit form of $f_{i \mu \nu s}$, identities
\beq
\sigma^{\mu \nu} \gamma_5 = -\frac{i}{2} \epsilon^{\mu \nu \lambda \sigma}
\sigma_{\lambda \sigma}
\eeq
to get rid of $\epsilon$-couplings and discarding terms proportional to
$q^2$ (on-shell photons), we write
\begin{eqnarray}
h^\mu_{\alpha \alpha}=i \bar{b}(p') \Bigl \{ p_s^\mu((ql)+2 \cfd (kq))-
l^\mu(p_sq)+q^\mu(2(p_sl)-4 \cfd (p_s k) + \dfd (p_s q) + \nonumber \\
+ 2 \cfd k^3 \frac{p_s q}{kq} ) -
k^\mu (2 \cfd (p_s q)) + i k^\mu ( 2 \cfd \frac{p_s q}{kq} \sab k_\alpha
q_\beta + \cfd \sab q_\alpha p_{s \beta}) \\
+ 2i (p_s q) \smr l_r
- 2i(ql) \smr p_{s \alpha} +2i (p_sl) \smr q_\alpha -2i \cfd  (p_sk)
\sma q_\alpha + 2i \cfd ((p_s q) \sma k_\alpha  \nonumber \\
- \frac{k^2}{kq} \sma q_\alpha)
- i \cfd (kq) \sma p_{s \alpha} - \sab q_\alpha l_\beta \smr p_{sr} \Bigr\}
(1 + \gamma_5) b(p) \nonumber
\end{eqnarray}

Using (\ref{de}) we obtain the contributions from class $III$. Finally,
class $IV$ gives the following

\begin{eqnarray}
T_{IV}=-\frac{i}{k^2} \frac{G_F}{\sqrt{2}} \frac{e Q}{8 \pi^2} V_{bi} V_{is}
F_2^\gamma (x) m_b g_s^2
\Bigl\{ \bar{s} (p_s)
\sigma_{\alpha \nu} q^\nu ( 1 + \gamma_5 ) \lambda_a b (p_b) \nonumber \\
\bar{u} (p_u)
\Bigl( \frac{2 \slash{p}_s}{2 p_sk+k^2} -
\frac{2 \slash{p}_b}{2 p_bk - k^2} \Bigr) \lambda_a u (p_u) -
\frac{1}{2 p_bk - k^2} \bar{s} (p_s) \sigma_{\alpha \nu} q^\nu ( 1- \gamma_5 )
\slash{k} \gamma_\mu \lambda_a b (p_b) \\
\bar{u} (p_u') \gamma^\mu \lambda_a u (p_u) +
\frac{1}{2 p_sk + k^2} \bar{s} (p_s) \gamma_\mu \slash{k} \sigma_{\alpha \nu}
q^\nu ( 1- \gamma_5 ) \lambda_a b (p_b) ~ \bar{u} (p_u') \gamma^\mu \lambda_a
u (p_u)
\Bigr\} \nonumber
\end{eqnarray}

Let us also mention that some range of momentum transferred to the
spectator gives a contribution to the decay rate
which is highly suppressed - in the ACCMM-model, for instance, decay rate
looks like
\beq
\Gamma=\int d^3p \phi^*_B(p') \Gamma_{qm} \phi_B(p)
\eeq
with $\vec{p}'=\vec{p}+\vec{k}$ and $\phi(p)=(1/p^3 \sqrt{\pi})
exp(-\bar{p}^2/2p_f^2)$. Note that this expression reduces to the
conventional ACCMM-averaging if the momentum flow between heavy
quark and spectator lines is discarded. It is clear that if the magnitude
of $3$-momentum transferred to the spectator line is large, the factor of
wave function suppression is large as well which results in
the exponential fall-off of the $\alpha_s$-spectrum as $E_\gamma \to 0$.
This is the reason why
non-spectator $\alpha_s$ corrections turn out to give much smaller impact on
the line shape of the spectrum than those at order $\alpha_s^2$,
in fact,
\beq
\frac{\Gamma_{mix}}{\Gamma_{sc}} \leq 1.0 \cdot 10^{-2}
\eeq

\subsection{Corrections of order $\alpha_s^2$.}

Let us now consider corrections of order $\alpha_s^2$. In fact they
can simply be obtained by squaring the sum of $T_i$ presented in the
previous section. We shall present only few of them emphasizing
their possible experimental signatures and highlighting some steps of
derivations.

{\bf Class I}. Diagrams of class $I$ give contributions to $\Gamma_{q}$
and have been calculated elsewhere \cite{quark}.
This contribution to the branching ratio $Br$ varies from $2.0 \cdot 10^{-4}$
to
$4.2 \cdot 10^{-4}$. The major uncertainties at present come from the
choice of the renormalization scale in the effective operators and
the KM elements,
and so this branching ratio can be predicted more accurately in the future
as the KM elements are more precisely measured.

{\bf Class II}. Diagrams of this class give a new contribution to
$\Gamma_{tot}$ which comes from the emission of the photon from the spectator
quark.

The matrix elements for diagrams ($6$) and ($7$) ( Fig. 1 ) can be rewritten as
\beq
T_{fi}=\frac{1}{k^2} H_\mu L_\alpha^\mu \epsilon^\alpha
\eeq
and are given by (\ref{t2}). Taking the hermitian conjugate and multiplying
by (\ref{t2}) gives
$|T_{fi}|^2$ for which we can write
\beq \label{master}
|T_{fi}|^2=\frac{C_\alpha^2 (4 \pi)^2 \alpha_s \alpha}{k^4} H_{\mu \nu}
L^{\mu \nu}_{\alpha \beta} W^{\alpha \beta}
\eeq
where
\begin{eqnarray}
W^{\alpha \beta}= \sum_{pol} \epsilon^{\alpha*} \epsilon^\beta =
-g^{\alpha \beta} \nonumber \\
H_{\mu \nu}=
Tr \Bigl\{\slash{p}_b~ \Gamma_\mu^a~ \slash{p}_s ~ \Gamma_\nu^b \Bigr\} \\
L_{\alpha \beta}^{\mu \nu a b} = Tr \Bigl\{\slash{p}_u' \Pi^{\mu a}_\alpha
\slash{p}_u \Pi^{\nu a}_\beta \Bigr\} \nonumber
\end{eqnarray}
where as before we set $m_s=0$ and $C^2_\alpha$ is a color factor.
To obtain $C^2_\alpha$ we have used the fact that $B$-meson is a
superposition of the color-singlet states, so
\beq
C^2_\alpha=\frac{1}{3} tr \Bigl(
\frac{\lambda_a}{2} \frac{\lambda_b}{2} \frac{\lambda_b}{2}
\frac{\lambda_a}{2} \Bigr) = \frac{16}{9}
\eeq
Evaluating the traces we obtain for $H_{\mu \nu}$:
\begin{equation}
H_{\mu \nu}=\frac{G_F^2}{32 \pi^3} \alpha_s
|V_{bi} V_{is}|^2 F_1^{g~2} k^4
\Bigl\{ p_{s \mu} p_{b \nu} + p_{s \nu} p_{b \mu} - (p_s p_b)g_{\mu \nu} +
i \epsilon_{\mu \nu \alpha \beta} p_s^\alpha p_b^\beta \Bigr\}
\end{equation}
where the last term vanishes upon contraction with
$L_{\alpha \beta}^{\mu \nu}$ for which we can write down (taking into account
contraction with $W_{\alpha \beta}$):
\begin{eqnarray}
L_{\alpha \alpha}^{\mu \nu}=Q_{sp}^2 (2(p_u p_1) (({p_u^\mu}'p_1^\nu)
+ {p_u^\nu}'p_u^\mu
- g^{\mu \nu}(p_u' p_u)) - p_1^2 ({p_u^\mu}' p_u^\nu + {p_u^\nu}'
p_u^\mu - \nonumber \\
g^{\mu \nu} (p_u' p_u))
+ 2 p_2^\mu p_2^\nu (p_u p_u') + g^{\mu \nu} (2 (p_u' p_2)(p_u p_2) -
p_2^2 (p_u p_u') + \\
p_2^2(p_u^\mu {p_u^\nu}' + p_u^\nu {p_u^\mu}')) -
2 {p_u^\mu}' p_2^\nu (p_u p_2) -2 p_u^\nu p_2^\mu (p_u' p_2)) \nonumber
\end{eqnarray}
for the diagrams ($6$) and ($7$). These diagrams might be interesting
from the experimental point of view
since they give rise to different contributions from charged and neutral
$B$-mesons (the amplitude is proportional to the charge of
spectator $Q_{sp}$), so one, in principle, can observe this charge
asymmetry
\beq
A=\frac{\Gamma(B^+ \to X_s \gamma)-\Gamma(B^0 \to X_s \gamma)}{\Gamma(B^+
\to X_s \gamma) + \Gamma(B^0 \to X_s \gamma)} \sim 0.3 \cdot 10^{-3}
\eeq
However the small size of our prediction means that it will be difficult
to make this measurement.

{\bf Class III}. Calculation of the diagrams of this class involves the
effective vertex (\ref{triangle}) for the sum of two diagrams - 8 and 9.
As before, the amplitude squared can be written in the form
\begin{eqnarray}
|T_{fi}|^2=\frac{C_\alpha^2}{k^4} H_{\mu \nu}^{\alpha \beta ab}
L^{\mu \nu}_{ab} W^{\alpha \beta} \nonumber \\
H_{\mu \nu ab}^{\alpha \beta}=Tr \Bigl\{\Gamma_\mu^{\alpha a}~
\slash{p}_s~ \Gamma_\nu^{\beta b}~ \slash{p}_b \Bigr\} \\
L^{\mu \nu}=
4 \Bigl\{ p_u^\mu {p_u^nu}' + {p_u^\mu}' p_u^nu - (p_u p_u')g^{\mu \nu}
\Bigr\} \nonumber
\end{eqnarray}
Expanding $\Gamma_\mu^{\alpha a}$ as $\Gamma_\mu^{\alpha a} = \sum_i
f^a_{i~\mu \alpha s} \gamma_s (1 + \gamma_5)$ we get
\begin{eqnarray}
H_{\mu \nu} = \sum_{i k} f_{i \mu \alpha s} f_{k \nu \alpha t}
Tr \Bigl\{ \gamma_s (1+\gamma_5) \slash{p}_s \gamma_t (1+\gamma_5)
\slash{p}_b \Bigr\} \nonumber \\
= 8 \sum_{i k} f_{i \alpha s \mu} f_{k \alpha t \nu}
\Bigl\{ p_{s s} p_{b t} + p_{s t} p_{b s} - (p_s p_b)g_{st} +
i \epsilon_{stnm} p_s^n p_b^m \Bigr\}
\end{eqnarray}
Thus, the problem is the calculation of $\sum_{i k} f_{i \alpha s \mu}
f_{k \alpha t \nu}$. This calculation, which requires the explicit form for
the $f$-functions and involves expansions of the products of
$\epsilon$-tensors, results in a quite lengthy final expression. However,
restricting our attention only to the case of real photons ($q^2=0$) and
taking advantage of the symmetry of $L^{\mu \nu}$ under
$\mu \leftrightarrow \nu$ (implying
further convolution with ``light'' tensor $L^{\mu \nu}$ ) simplify
the result enormously since
there are only four combinations of $f$'s that
give non-vanishing contributions to the problem of interest. They are
\begin{eqnarray}
f_{1~ \mu s \alpha} f_{1 \nu s \beta}= - (l^2(g_{\mu \nu} g_{\alpha \beta} -
g_{\alpha \nu} g_{\mu \beta} ) + l_\mu l_\beta g_{\alpha \nu} +
l_\nu l_\alpha g_{\mu \beta} -
l_\alpha l_\beta g_{\mu \nu} - l_\mu l_\nu g_{st})  \nonumber \\
f_{1~ \mu s \alpha} f_{2~ \nu s \beta} +
f_{2~ \mu s \alpha} f_{2~ \nu s \beta} +
f_{2~ \mu s \alpha} f_{1~ \nu s \beta} =
-\Bigl(\frac{{\Delta i_{23}}^2}{(kq)^2}k^2 +
2 \frac{\Delta i_{23} \Delta i_{6}}{kq} \Bigr)
\epsilon_{\rho \sigma \mu \alpha} \epsilon_{mn \nu \beta} k^\rho
q^\sigma k^m q^n \nonumber \\
f_{3~ \mu s \alpha} f_{3~ \nu s \beta}=-q_\mu q_\nu \Bigl( \frac{\Delta i_{26}}
{kq} \Bigr)^2 \Bigl\{ (kq)(q_\alpha k_\beta +q_\beta k_\alpha -
(kq)g_{\alpha \beta}) - q_\alpha q_\beta k^2 \Bigr\}  \\
f_{3~ \mu s \alpha} f_{1~ \nu s \beta} +
f_{1~ \mu s \alpha} f_{3~ \nu s \beta}
= \frac{\Delta i_{26}}{kq} \Bigl\{
q_\mu q_\nu ( - 2(lk) g_{\alpha \beta} + k_\alpha l_\beta + l_\alpha
k_\beta) - q_\alpha q_\nu ( k_\mu l_\beta - \nonumber \\
(lk) g_{\mu \beta}) -
q_\beta q_\mu (k_\nu l_\alpha - (lk) g_{\alpha \nu}) +
g_{\alpha \beta} (k_\mu q_\nu + q_\mu k_\nu)(lq) -
(lq)(k_\alpha q_\nu g_{\mu \beta} + k_\beta q_\mu g_{\alpha \nu})
\Bigr\}  \nonumber
\end{eqnarray}
Combining all of the terms together and
contracting $H_{\mu \nu}^{\alpha \alpha}$ to
$L^{\mu \nu}$ one obtains the contribution of the class III diagrams to
the $\Gamma_{tot}$ which is to be computed numerically. The
resulting analytical expression for the amplitude is too long to be
presented here.

{\bf Class IV}. Finally for the class of diagrams responsible for the
$b$($s$)-spectator hard gluon exchange, using the same
technique  as above, we write for $L$ and $H$ tensors
\begin{eqnarray}
L^{\mu \nu}_{ab}=4 \Bigl\{ p_u^\mu {p'}_u^\nu + {p'}_u^\mu p_u^\nu -
(p_u p_u')g^{\mu \nu} \Bigr\} \nonumber \\
H_{\mu \nu ab}^{\alpha \alpha}=Tr \Bigl\{\slash{p}_s~ \Pi_\mu^a~ \slash{p}_b~
\Pi_\nu^b \Bigr\}
\end{eqnarray}
Here the ``heavy'' current is
\beq
H^{\alpha a}_\mu= \bigl( \bar{s}(p_s) \Gamma^{\alpha}
\frac{1}{\slash{p}_1-m_b} \gamma_\mu
b(p_b) + \bar{s}(p_s) \gamma_\mu \frac{1}{\slash{p}_2} \Gamma^{\alpha}
b(p_b)) = \bar{s}(p_s) \Pi^{\alpha a}_\mu b(p_b)
\eeq
and $\Gamma_\mu$ is given by (\ref{onshell}).
Evaluating the traces and making use of the facts that $L^{\mu \nu}_{ab}$
is symmetric under $\mu \leftrightarrow \nu $ and $q^2=0$, we obtain
\begin{eqnarray}
H_{\mu \nu}^{\alpha \alpha}=
\frac{G_F^2}{2 \pi} m_b^2 F_2^{\gamma~ 2} (x) \alpha \alpha_s^2 Q_{int}^2
|V_{bi}V_{is}|^2 C_\alpha^2
\Bigl[ \frac{32}{k_1^4(p_1^2-m_b^2)^2} \Bigl((p_1^2-m_b^2)
\nonumber \\
(p_b^\nu q^\mu + p_b^\mu q^\nu -
(p_b q)g^{\mu \nu}) - 2(pq)(p_1^\mu p_b^\nu + p_1^\nu p_b^\mu -
(p_1 p_b) g^{\mu \nu}) \Bigr) + \\
\frac{32}{k_2^4 p_2^4} (p_b q) \Bigl( p_2^2 (p_s^\mu q^\nu + p_s^\nu
q^\mu - (p_s q) g^{\mu \nu} ) - 2 (p_2 q)(p_s^\mu p_2^\nu +
p_s^\nu p_2^\mu - (p_s p_2) g^{\mu \nu} ) \Bigr) \Bigr] \nonumber
\end{eqnarray}
with $p_1=p_s+q$ and $p_2=p_b-q$, here $Q_{int}$ is a charge of the
internal quarks. Contracting
$H_{\mu \nu ab}^{\alpha \alpha}$ with $L^{\mu \nu}_{ab}$
gives the expression for the diagrams of class $IV$.
In the final calculation of the magnitude of the spectator effects, the
dominant contribution comes from the class $IV$ diagrams \cite{milana}.

In evaluating the matrix elements, we have taken the $b$-quark spinors to be
non-relativistic. The interferences between the different classes of diagrams
involving the spectator quark have also been taken into account, although
we have not displayed the (lengthy) formulas explicitly above. Some of the
complicated sums over many Dirac matrices were evaluated numerically
with the help of $Mathematica^{TM}$.

Substituting $|T_{fi}|^2$ and (\ref{phres}) into (\ref{rate}),  and
performing numerical $5$-dimensional Monte-Carlo integration over
$\phi_s,\phi_q,\theta_q,\theta_s,p$ we obtain
photon energy distribution presented in Fig.3.
Integrating over $x = 2 \omega / M_b$ gives for the decay rate:
\begin{eqnarray}
\delta \Gamma_{tot} \sim 5.3 \cdot 10^{-18}~GeV \\
\delta Br \sim 1.5 \cdot 10^{-5}
\end{eqnarray}
Given the recent measured by $CLEO$ value for $Br(B \to X_s \gamma) =
2.32 \cdot 10^{-4}$ \cite{cleo}
we conclude that spectator corrections give an overall effect of about
$5 \% $ on the inclusive decay rate $\Gamma(B \to X_s \gamma)$.

\section{Summary}
The diagrams which involve the spectator quark are in general non-local and
involve momentum transfers to the spectator which are comparable to the energy
released in the decay. This is a quite different configuration from the
$b \to s \gamma$ transition where the spectator quark is nearly at rest.
Despite the inclusion of the bound state momentum of the spectator, the
interference of the two classes of diagrams turns out to be very small.
We have evaluated the remaining effects of the spectator diagrams, which
then amount to effects at order $\alpha_s^2$.

Our results indicate that the diagrams involving the spectator
quark do not play a large role in $B \to X_s \gamma$, which
implies that the $b \to s \gamma$ reaction is
the dominant contribution to the inclusive decay.

{\bf \large{Acknowledgments.}}
One of us (A.P.) would like to thank S. Nikolaev for useful conversations
regarding Monte-Carlo integration. This work has been supported in part by
the US National Science Foundation.

{\bf \large{Appendix. Calculation of the Phase Space.}}
Since we are calculating $QCD$-corrections associated with bound-state
effects it is convenient to work in the rest frame of $B$-meson.
Let us adopt ACCMM-model for use in this particular frame.
In this coordinate system (the center of mass frame for the system
``heavy quark-spectator''):
\begin{eqnarray}
\vec{p}_u'+\vec{p}_s=0=\vec{p}_u+\vec{p}_b \\
E_b+E_u=M_B=E_u'+E_s+\omega
\end{eqnarray}
We treat the spectator as a massless particle,so
\beq
E_u=p_u=p,~~~E_b=\sqrt{m_b^2+p^2}
\eeq
As usual in the ACCMM model the effective $b$ quark mass is momentum
dependent and is given by
$m_b^2=M_B(M_B - 2 p)$. Also
\beq \label{eu}
E_s=\sqrt{\vec{p}_s^2}=\sqrt{(\vec{p'}_u + \vec{q})^2}=
\sqrt{{E'}_u^2+\omega^2+2{E'}_u \omega cos(\theta'_{uq})}
\eeq
Let us note that in this frame the decay products
$u',s,q$ lay in the same plane,
thus the angle parameters are
$\theta'_{uq}$-the angle between the directions of $3$-momenta of $s$-quark
and photon and the relative angle between $3$-momenta of the initial
(``spectator'') $u$-quark and normal to the decay plane.
Let us recall that
\beq
\delta^4(p_f-p_i)=\delta(E_f-E_i) \delta^3(\vec{p}_f-\vec{p}_i)
\eeq
Therefore, (\ref{phase}) can be integrated over $\vec{p}_u'$ to get
\beq
d \Phi \sim \frac{d^3 q}{\omega} \frac{d^3 p'_u}{E_u'E_s} \delta( M_B -
E_u'-E_s-\omega)
\eeq
with $M_B^2=(E_b+E_u)^2$. Next, evaluating the $\vec{p'}_u$-integral
(keeping in mind that $E_s$ is given by (\ref{eu})) yields
\begin{equation}
d \Phi \sim \frac{d^3 q}{\omega} \frac{d \Omega'_u {E'}_u dE'_u}{E_s}
\delta(M_B-E_u'-E_s-\omega)
\end{equation}
Making use of a familiar $\delta$-function relation
\beq
\delta(f(x))=\sum_{x=x_0} \frac{1}{|f'(x_0)|} \delta(x-x_0)
\eeq
to perform integration over $E'_u$ we arrive at the equation for
the roots of $E'_u$:
\beq
M_B-E'_u- \sqrt{{E'}_u^2+\omega^2+2E'_u \omega cos \theta'_{uq} } - \omega=0
\eeq
which can be solved to get
\beq
E'_{u_0}=\frac{M_B(M_B-2 \omega)}{2(M_B - \omega(1-cos \theta'_{uq}))}.
\eeq
This gives for the $\delta$-function:
\beq
\delta(E_f-E_i)=\frac{E_s(E'_{u_0})}{E'_{u_0} + E_s(E'_{u_0}) +
\omega cos' \theta_{uq} } \delta(E'_u-E'_{u_0})
\eeq
Hence, the phase volume is
\beq \label{phres}
d \Phi \sim \omega d \omega d \Omega_q d \Omega'_u \frac{M_B(M_B - 2 \omega)}
{2(M_B - \omega( 1 - cos \theta'_{uq} ) )^2 }
\eeq
with $ d \Omega_i = d cos \theta_q d \phi$.

\begin{figure}[t]
\centering
\leavevmode
\epsfysize=600pt
\centerline{
\epsfbox{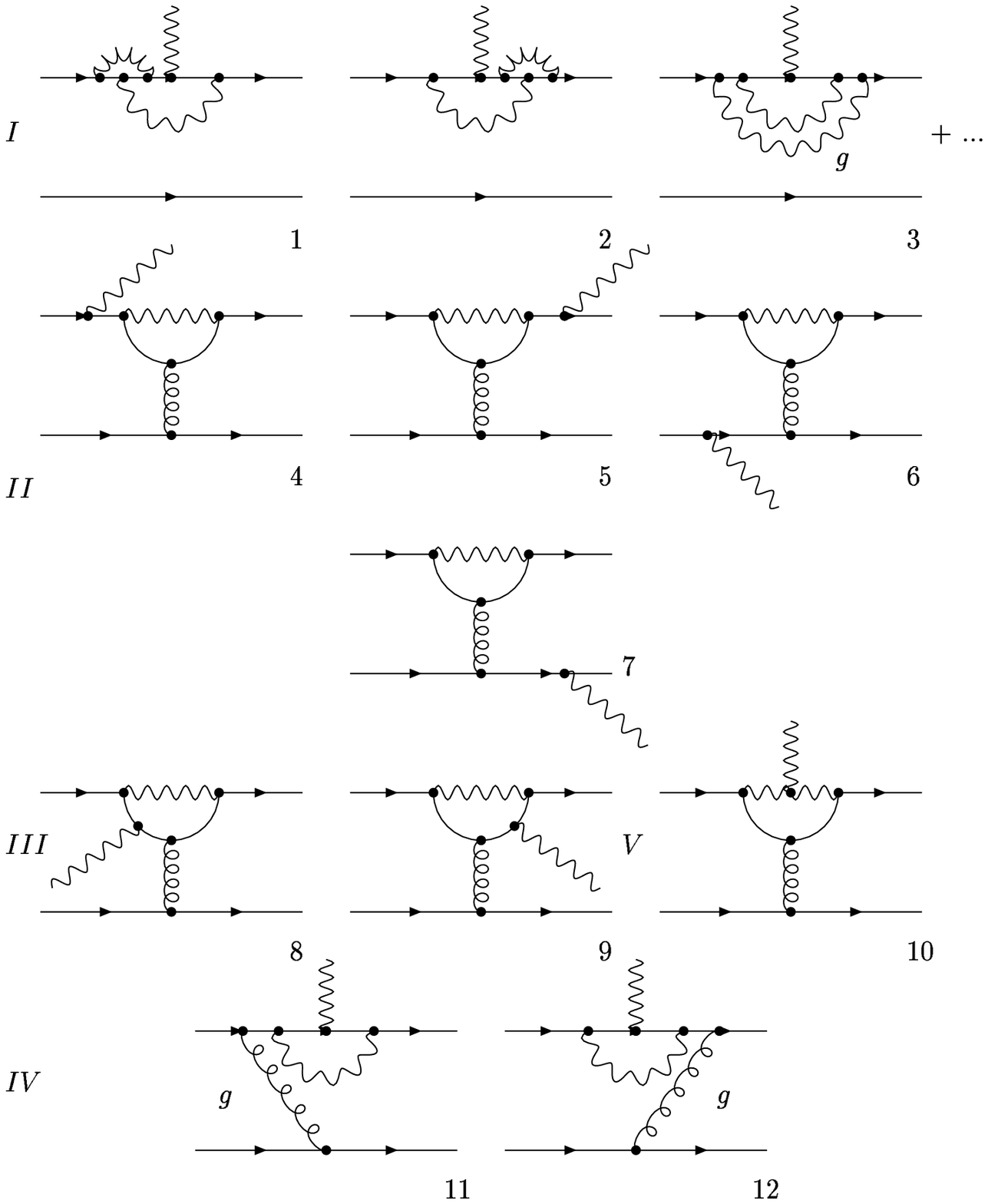}}
\caption{Set of the relevant diagrams.}
\end{figure}

\begin{figure}[t]
\centering
\leavevmode
\epsfysize=600pt
\centerline{
\epsfbox{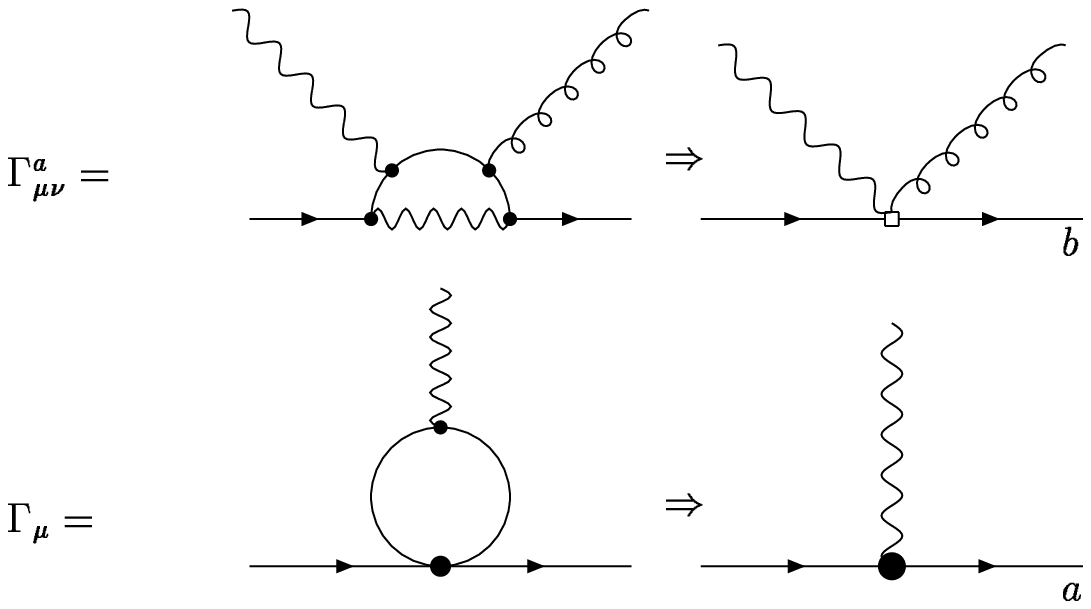}}
\caption{Effective vertices.}
\end{figure}

\begin{figure}
\centerline{
\epsfbox{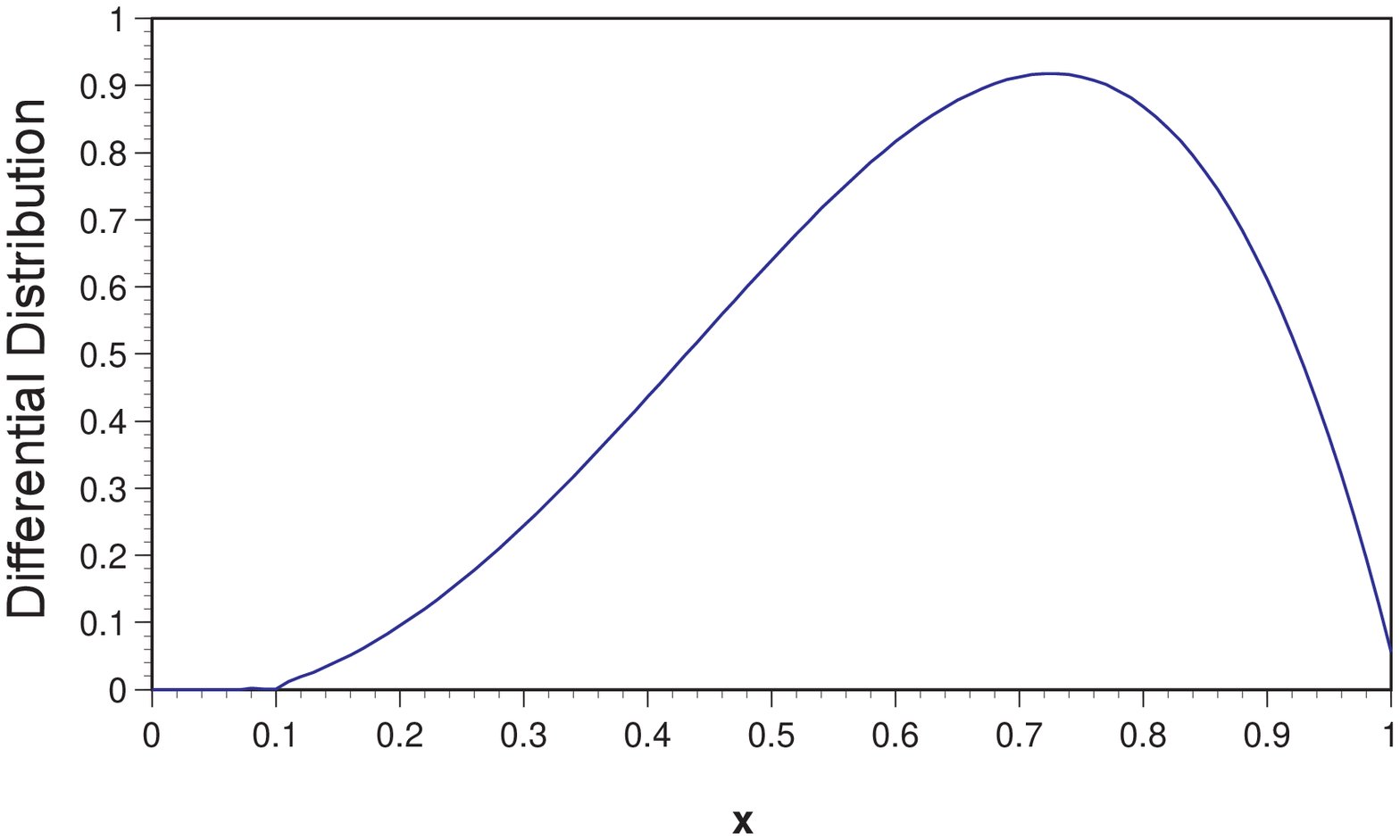}}
\caption{Spectrum for $B \to X_s \gamma$.}
\end{figure}


\begin{thebibliography}{99}
\bibitem{hewett}{J. Hewett, SLAC-PUB-95-6782 [hep-ph/9505247] (Talk
presented at International Workshop on B Physics, Nagoya, Japan, Oct.1994.)}
\bibitem{cleo}{CLEO collaboration, CLEO 94-25.}
\bibitem{quark}{S. Bertolini, F. Borzumati, A. Masiero,
			   Phys. Rev. Lett. 59 (1987) 180. \\
		M. Misiak, Nucl. Phys. {\bf B393} (1993) 23. \\
B. Grinstein, R. Springer, M.B. Wise, Nucl. Phys. {\bf B339} (1990) 269. \\
		A. Ali, C. Greub, Z. Phys. {\bf C49} (1990) 431. \\
A.F. Falk, M. Luke, M.J. Savage, Phys. Rev {\bf D49} (1994)3367. \\
	A.J. Buras et. al. Nucl. Phys. {\bf B424} (1994) 374. \\
J.M. Soares, TRI-PP-95-06, hep-ph/9503285. }
\bibitem{accmm}{G. Altarelli et al., Nucl. Phys. {\bf B208} (1982)365.}
\bibitem{shifman}{M.A. Shifman, A.I. Vainshtein, V.I. Zakharov,
			Phys.Rev. {\bf D18} (1978) 2583, \\
		  T. Imami, C.S. Lim, Progr. of Theor. Phys.
			{\bf 65} (1981) 297.}
\bibitem{deshpande}{N.G. Deshpande, J. Trampetic, Phys.Rev. {\bf D41}
				(1990) 895.}
\bibitem{rosen}{L. Rosenberg, Phys.Rev. {\bf D129} (1963) 2786.}
\bibitem{wyler}{D. Wyler, H.Simma, Nucl.Phys. {\bf B344} (1990) 283, \\
		C. Greub, D. Wyler, H. Simma, Nucl. Phys. {\bf B434}
			(1995) 39.}
\bibitem{milana}{C.E. Carlson, J. Milana, hep-ph/9405344, \\
		 J. Milana, hep-ph/9503376.}
\bibitem{dikeman}{R.D. Dikeman, M. Shifman, N.G. Uraltsev,
			hep-ph/9505397, \\
       		  A. Ali, J. Phys. {\bf G18} (1992) 1605, \\
	          A. Ali, DESY 95-157 [hep-ph/9508335] (review).}
\bibitem{resonance}{A. Ali, T. Ohl, T. Mannel, Phys. Lett. {\bf B298}
			(1993) 195.}
\end{thebibliography}
\end{document}